\documentstyle{article}
\input{epsf}
\begin{document}
\title{{\bf The Fractal Properties of the Source and BEC}
\footnote{Talk presented by O.V.Utyuzh at the $12^{th}$
{\it Indian Summer School "Relativistic Ion Physics"}
held in Prague, Czech Republic, 30 August - 3 September
1999, to be published in {\sl Czech J. Phys.} }}
\author{O.V.Utyuzh$^{1}$\footnote{e-mail: utyuzh@fuw.edu.pl},
        G.Wilk$^{1}$\thanks{e-mail:wilk@fuw.edu.pl} and 
        Z.W\l odarczyk$^{2}$\thanks{e-mail:wlod@pu.kielce.pl}\\
        [2ex]  
 $^1${\it The Andrzej So\l tan Institute for Nuclear Studies,}\\
     {\it Ho\.za 69; 00-689 Warsaw, Poland}\\
     [1ex]
 $^2${\it Institute of Physics, Pedagogical University,}\\
     {\it Konopnickiej 15; 25-405 Kielce, Poland}\\  
 }
\date{\today}
\maketitle 

\begin{abstract}
Using simple space-time implementation of the random cascade model we 
investigate numerically influence of the possible fractal structure
of the emitting source on Bose-Einstein correlations between
identical particles. The results are then discussed in terms of the
non-extensive Tsallis statistics. 
\end{abstract}

\section{Formulation of the problem}
\subsection{Introduction}
Two features seen in the analysis of multiparticle spectra of
secondaries produced in high energy collision processes are of
particular interest: 
$(i)$ intermittent behaviour observed in analysis of factorial
moments and $(ii)$ Bose-Einstein correlations (BEC) observed between
identical particles. Whereas the former indicates a possible (multi)
fractal structure of the production process (in the momentum space)
\cite{CAS} the latter provides us with knowledge on the space-time aspects
of production processes \cite{BEC}.
It was argued \cite{B} that these features are compatible with each
other only when: 
$(i)$ either the shape of interaction region is regular but its size
fluctuates from event to event according to some power-like scaling
law or 
$(ii)$ the interaction region itself is a self-similar fractal (in
coordinate space) extending over a very large volume. Although there
exists a vast literature on the (multi)fractality 
in momentum space \cite{HWA} its space-time aspects are not yet fully
recognized with \cite{B} remaining so far the only representative
investigations in this field. We shall present here numerical
analysis of a particle production model possessing both the momentum
and coordinate space fractalities \cite{UWW} extending therefore
ideas discussed in \cite{B} to a more realistic scenario.

\subsection{Cascade model used}
As a model we shall choose a simple self-similar cascade process
of the type discussed in \cite{SS} but developed further to meet our
demands. The following point must be stressed. Every cascade model is
expected to lead automatically to intermittent behaviour of momentum
spectra of observed particles \cite{ZAL}. Although this is true for
models based on random multiplicative processes in observed variables
(like energy, rapidity or azimuthal angle), this is not necessarily
the case for multiplicative processes in variables which are not
directly mesurable but which are, nevertheless, of great dynamical
importance (like masses $M_i$ of intermediate objects in a cascade
process considered here). In a purely mathematical case, where
cascade process proceeds {\it ad infinitum}, one eventually always
arrives at some fractal picture of the production process. However,
both the finite masses $\mu$ of produced secondaries and the limited
energy $M$ stored originally in the emitting source prevent the full
development of such fractal structure \cite{CARR}. One must be
therefore satisfied with only limited and indirect presence of such
structure. This applies also to the analysis presented here.

In our model some initial mass $M$ ``decays'' into two masses, $M\,  
\longrightarrow \, M_1\, +\, M_2$ with $M_{1,2} = k_{1,2}\cdot M$ and
$k_1 + k_2 < 1$ (i.e., a part of $M$ equal to $(1- k_1 - k_2) M$
is transformed to kinetic energies of  the decay products $M_{1,2}$).
The process repeats itself until $M_{1,2} \ge \mu$ ($\mu$ being the
mass of the produced particles) with successive branchings occuring
sequentially and independently from each other, and with {\it a
priori} different values of $k_{1,2}$ at each branching, but with
energy-momentum conservation imposed at each step. For different
choices of dimensionality $D$ of cascade process, $D=1$ (linear) or
$D=3$ (isotropic), and for different (mostly random) choices of decay
parameters $k_{1,2}$ at each vertex, we are covering a variety of
different possible production schemes, ranging from one-dimensional
strings to isotropic three-dimensional thermal-like fireballs. For
our purpose of investigation of connections between BEC and
space-time fractality of the source we have extended this (momentum
space) cascade also to the space-time and we have added to it a kind
of BEC ``afterburner'' along the lines advocated recently in
\cite{GEHW}.    

What concerns space-time development we model it by introducing a
fictitious finite life time $t$ for each vertex mass $M_i$,
distributed according to some prescribed distribution law given by
\begin{equation}
\Gamma (t) \, =\, \frac{2-q}{\tau}\cdot
                      \left[ 1 - (1 - q) \frac{t}{\tau}
                      \right]^{\frac{1}{1-q}}
   \qquad
   \stackrel{|q - 1| \rightarrow 0}{\Longrightarrow}
   \qquad
 \Gamma (t) \, =\, \frac{1}{\tau}\cdot
                   \exp{\left[ - \frac{t}{\tau}\right]}
               .
               \label{eq:Gamma}
\end{equation}
This procedure is purely classical, i.e., intermediate masses $M_i$
are not treated as resonances (as was done in \cite{PIS}) but are
regarded to be stable clusters with masses given by the corresponding
values of decay parameters $k_{1,2}$ and with velocities $\vec{\beta}
= \vec{P}_{1,2}/E_{1,2}$ ($(E_{1,2};\vec{P}_{1,2})$ are the
corresponding energy-momenta of decay products calculated in each
vertex in the rest frame of the parent mass). The energy-momentum and
charges are strictly conserved in each vertex separately. The form of
$\Gamma$ used in (\ref{eq:Gamma}) allows to account for the possible
fluctuations of the evolution parameter $\tau$ \cite{WWQ} and finds
its justification in the Tsallis statistics, to which we shall return
later when discussing our results \cite{T}.

\subsection{Bose-Einstein correlations}

Our main goal is the investigation of the BEC, in particular whether
these correlations show indeed some special features which could be 
attributed to the branchings and to their space-time and momentum
space structure. We are therefore interested in two particle
correlation function
\begin{equation}
C_2(Q=|p_i - p_j| )\, =\, 
             \frac{d N(p_i,p_j)}{dN(p_i)\, dN(p_j)} . \label{eq:C2}
\end{equation}
To calculate it we have decided to use the ideas of the BEC
``afterburners'' advocated recently in \cite{GEHW}. Such step is
necessary because cascade {\it per se} do not show bosonic bunching
in momenta (as is the case in models where Bose statistics is
incorporated from the beginning, like \cite{OMT}; however, we
cannot follow this strategy here). Because we are interested in some
possible systematics of results rather then in particular values of
the ``radius'' $R$ and ``coherence'' $\lambda$ parameters
characterizing source $M$, we have chosen the simplest, classical
version of such afterburner. After generating a set of
$i=1,\dots,N_l$ particles for the $l^{th}$ event we choose all pairs
of the same sign and endow them with the weight factors of the form  
\begin{equation}
C = 1 + \cos\left[ \left(r_i - r_j)(p_i - p_j\right)\right]
\label{eq:C} 
\end{equation}
where $r_i = (t_i,\vec{r}_i)$ and $p_i = (E_i,\vec{p}_i)$ for a given
particle. The signs are connected with charges which each cascade
vertex is endowed with using simple rules:
$\left\{0\right\}  \rightarrow  \left\{+\right\}+\left\{-\right\}$, 
$\left\{+\right\}  \rightarrow  \left\{+\right\}+\left\{0\right\}$
and 
$\left\{-\right\}  \rightarrow  \left\{0\right\}+\left\{-\right\}$.

\section{Results}
Although it is straightforward to get our cascade model in the form
of the Monte Carlo code, the main features of $D=1$ case can be also
demonstrated analytically. For example, in the limiting cases of
totally symmetric cascades (where for all vertices $k_{1,2}=k$), 
in which amount of energy allocated to the production is maximal, one
gets the following multiplicity of produced particles:
\begin{equation}
N_{s}\, =\, 2^{L_{max}}\, =\, 
           \left( \frac{M}{\mu}\right)^{d_F},
           \qquad d_F\, =\, \frac{\ln 2}{\ln\frac{1}{k}} .
            \label{eq:NSYM}
\end{equation}
It is entirely given by the length of the cascade, $L_{max} =
\ln (M/\mu) / \ln(1/k)$, $\mu = \sqrt{m^2_0 + \langle p_T\rangle^2}$.
The exponent $d_F$ is formaly nothing but a generalized (fractal)
dimension of the fractal structure in phase space formed by our
cascade. Notice the characteristic power-like behaviour of $N_{s}(M)$
in (\ref{eq:NSYM}) which is normally atributed to thermal models. For
example, for $k=1/4$ one has $N_{s} \sim M^{1/2}$, which in thermal
models would correspond to the ideal gas equation of state with
velocity of sound $c_0 = 1/\sqrt{3}$ \cite{UWW}. In the opposite
limiting case of maximally asymmetric cascades, $M \rightarrow \mu +
M_1$ (where $k_1 = \mu/M$ and $k_2= k$) in which the amount of
kinetic energy allocated to the produced secondaries is maximal, the
corresponding multiplicity is equal 
\begin{equation}
N_{a}\, =\, 1\, +\, L_{max} \, =\, 
     1\, +\, \frac{1}{\ln\frac{1}{k}}\cdot
                      \ln \frac{M}{\mu} . \label{eq:NASYM}
\end{equation}                      
The dependence on $L_{max}$ is now linear (i.e., dependence on the
energy is logarithmic). The important feature, which turns out to be
valid also in general, is the observed scaling in the ratio of the
available mass of the source $M$ and the mass of produced
secondaries: $M/\mu$.

The $D=3$ case differs only in that the decay products can flow
in all possible directions, which are chosen randomly from the
isotropic angular distribution. To allow for some nonzero transverse
momentum in $D=1$, we are using the transverse mass $\mu = 0.3$ GeV.
For $D=3$ cascade $\mu$ is instead put equal to the pion mass, $\mu =
0.14$ GeV. All decays are described in the rest frame of the
corresponding parent mass $M_i$ in a given vertex. To get the final
distributions one has to perform a necessary number of Lorentz
transformations to the rest frame of the initial source mass $M$. As
an output we are getting in each run (event) a number $N_j$ of
secondaries of mass $\mu$ with energy-momenta $(E_j;\vec{P}_j)_i$ and
birth space-time coordinates $\left(t_j; \vec{r}_j\right)_i, ~i=1,
\dots, N_j$ (i.e., coordinates of the last branching). Results
presented in Figs. 1 and 2 are obtained from $50000$ such events.
Decay parameters $k_{1,2}$ were chosen randomly from a triangle
distribution $P(k) = (1 - k)$ (leading to a commonly accepted energy
behaviour of $N(M) \sim M^{0.4\div0.5}$, as discussed above). For
more detailed presentation of rapidity and multiplicity distributions
and demonstration of intermittent behaviour of factorial moments see
\cite{UWW}.   

Fig. 1 shows densities $\rho(r)$ of points of production and
correlation function $C_2(Q)$ as defined in (\ref{eq:C}) for $D=1$
and $3$ dimensional cascades originated from masses $M=10,~40$ and
$100$ GeV. The evolution parameter is set equal $\tau = 0.2$ fm 
(in \cite{UWW} we discuss also $\tau \sim 1/M$ case). The decay
function $\Gamma$ is taken exponential (i.e., $q=1$). We observe, as
expected in \cite{B}, a power-like behaviour of cascading source:
\begin{equation}
\rho(r)\, \sim\,  \left(\frac{1}{r}\right)^L 
\qquad r > r_0 , \label{eq:scaling}
\end{equation}
{\it but only} for $r > r_0$, i.e., for radii larger then some (not
sharply defined) radius $r_0$, value of which depends on all
parameters used: mass $M$ of the source, dimensionality $D$ and
evolution parameter $\tau$ of the cascade. Below $r_0$ the $\rho(r)$
is considerably bended, remaining almost flat for $D=1$ cascades. For
the limiting case of $M=100$ GeV the corresponding values of
parameter $L$ vary from $L=1.89$ for one $D=1$ cascades to $L=2.78$
for $D=3$ cascades. The shapes of $\rho(r)$ scale in the ratio
$M/\mu$ in the same way as the multiplicities discussed before (the
same remains also true for rapidity and multiplicity distributions,
cf. \cite{UWW}). One can summarize this part by saying that
power-like behaviour \cite{B} sets in (albeit only approximately)
only for long cascades (large values of $M/\mu$). It remains
therefore  to be checked whether (and to what extend) such conditions
are indeed met in the usual hadronic processes. 

The corresponding BEC functions $C_2$ show a substantial differences
between $D=1$ and $D=3$ dimensional cascades, both in their widths
and shapes. Whereas the former are more exponential-like (except,
perhaps, for small masses $M$) the latter are more gaussian-like with
a noticeably tendence to flattening out at very small values of $Q$
observed for small masses $M$. Also values of intercepts, $C_2(Q=0)$,
are noticeable lower for $D=3$ cascades. The length of the cascade
(i.e., the radius of the production region, cf. discussion of density
$\rho$ before) dictates the width of $C_2(Q)$. However, the
$M/\mu$ scaling observed before in shapes of source functions is lost
here. This is because $C_2$ depends on the differences of the momenta
$p=\mu\, \cosh y$, which do not scale in $M/\mu$. The flattening
mentioned above for $D=3$ cascades are the most distinctive signature
of the fractal structure combined with $D=3$ dimensionality of the
cascade. The correlations of the position-momentum type existing here
as in all flow phenomena are, in the case of $D=3$ cascades, not
necessarily vanishing for very small differences in positions or
momenta between particles under consideration. The reason is that our
space-time structure of the process can have in $D=3$ a kind of
``holes'', i.e., regions in which the number of produced particles is
very small. This is perhaps the most characteristic observation for
fractal (i.e., cascade) processes of the type considered here. This
feature seems to be more pronounced for diluted cascades corresponding
to $q<1$ case discussed below. 

Fig. 2 displays the same quantities but this time calculated for
$M=40$ GeV and for different values of parameter $q$ in the decay
function $\Gamma$ defined in (\ref{eq:Gamma}). This function is
written there in form of Tsallis distribution \cite{T}, which allows
to account for a variety of possible influences caused by, for
example, long-range 
correlations, memory  effects or for the possible additional fractal
structure present in the production process. They all result in a
non-extensitivity of some thermodynamical variables (like entropy)
with $|1-q|$ being  the measure of this non-extensitivity. In
practical terms of interest here, for $q < 1$ the tail of distribution
(\ref{eq:Gamma}) is depleted and its range is limited to $t\in (0,
\tau/(1 - q))$ whereas for $q>1$ it is enhanced in respect to the
standard exponential decay law (and its range is $t\in (0,\infty)$)
\cite{T}. In other words, one can account in this way for both more
diluted (for $q<1$) and more condensed (for $q>1$) space-time
structure of the developing cascades. Such distributions are
ubiquitous in numerous phenomena and they are founded in the, so 
called, Tsallis non-extensive thermostatistics \cite{T}  generalizing
the conventional Boltzmann-Gibbs one (which in this notation
corresponds to the $q = 1$ case). It has found also applications in
high energy and nuclear physics (cf. \cite{WWQ} for references). Its
effect on the cascades investigated here is, as can be seen from Fig.
2, in that it mimics (to some extent) the changes atributed in Fig. 1
to different energies (making cascade effectively shorter for $q=0.8$
and longer for $q=1.2$). The results of Fig. 2 (taken for $M=40$ GeV)
should be then compared with those of Fig. 1 for $M=10$ and $M=100$
GeV. They demonstrate that effects of longer or shorter cascades in
momentum space (as given by different $M$ in Fig. 1) is similar to
effects of the more or less condensed cascades in the position space
as given by $q$ here. This fact should be always kept in mind in such
analysis as ours. 

\section{Conclusions}
We conclude that BEC are, indeed, substantially influenced by the
fact that the production process is of the cascade type (both in
momentum and space-time) as was anticipated in \cite{B}, although
probably not to the extent expected (which, however, has not been
quantified there). In practical applications (fitting of experimental
data) there are many points which need further clarification. The
most important is the fact that data are usually collected for a
range of masses $M$ and among directly produced particles are also
resonances. This will directly affect lengths of the cascades,
and through them the final results for $C_2$. Selecting events with
similar masses $M$ should allow to check whether in such processes
$q=1$ or not. The importance of this finding is in that $q<1$ would
signal a non-stochastic development of the cascade, whereas $q>1$
would indicate that, as has been discussed in \cite{WWQ}, parameter
$\tau$ fluctuates with relative variance 
\begin{equation}
\omega \, =\, \frac{ \left\langle \left( \frac{1}{\tau} \right)^2
                     \right\rangle \, -\, 
                     \left\langle \frac{1}{\tau} \right\rangle^2 }
                   { \left\langle \frac{1}{\tau} \right\rangle^2 }
                   \, =\, q - 1 . \label{eq:OM}
\end{equation}
Such fluctuations are changing exponential behaviour of $\Gamma$ in
(\ref{eq:Gamma}) to a power-like distribution with enhanced tail.\\

\noindent {\bf Acknowledgement:} O.V.Utyuzh is very grateful to
organizers of the $12^{th}$ {\it Indian Summer School "Relativistic 
Heavy Ion Physics"} held in Prague, Czech Republic, $30$ August - $3$
September 1999, for financial support and warm hospitality extended
to him during the conference. \\

\newpage

{\bf Figure Captions}

\vspace{0.5cm}

\noindent Fig. 1. Density distribution of the production points 
$\rho(r)$ (left panels) and the corresponding $C_2(Q=|p_i - p_j|)$
(right panels) for one-dimensional (upper panels) and
three-dimensional (lower panels) cascades. Each panel shows results
for $M = 10,~40$ and $100$ GeV masses of the source. Time evolution
parameter is $\tau = 0.2$ fm and nonextensitivity parameter $q=1$.

\begin{figure}[h]
\begin{picture}(0,430)
\epsfxsize=10cm
\hspace{15mm} \epsffile{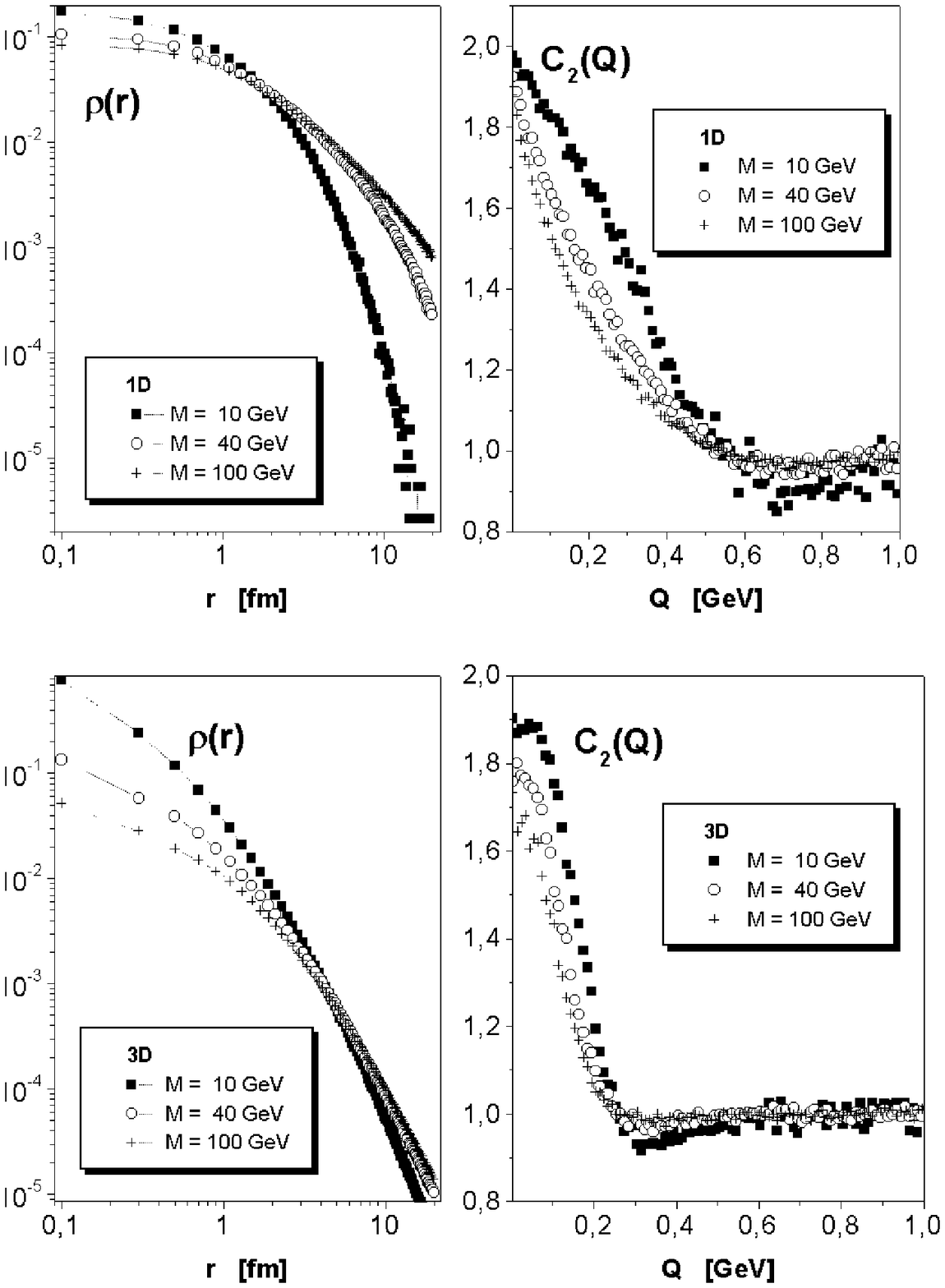}
\end{picture}
\end{figure}

\vspace{0.5cm}

\noindent Fig. 2. The same as in Fig. 1 except that each panel shows
results for mass of the source $M=40$ GeV and for three different
values of the nonextensitivity parameter $q=0.8,~1.0$ and $1.2$.

\begin{figure}[h]
\begin{picture}(0,450)
\epsfxsize=10cm
\hspace{15mm} \epsffile{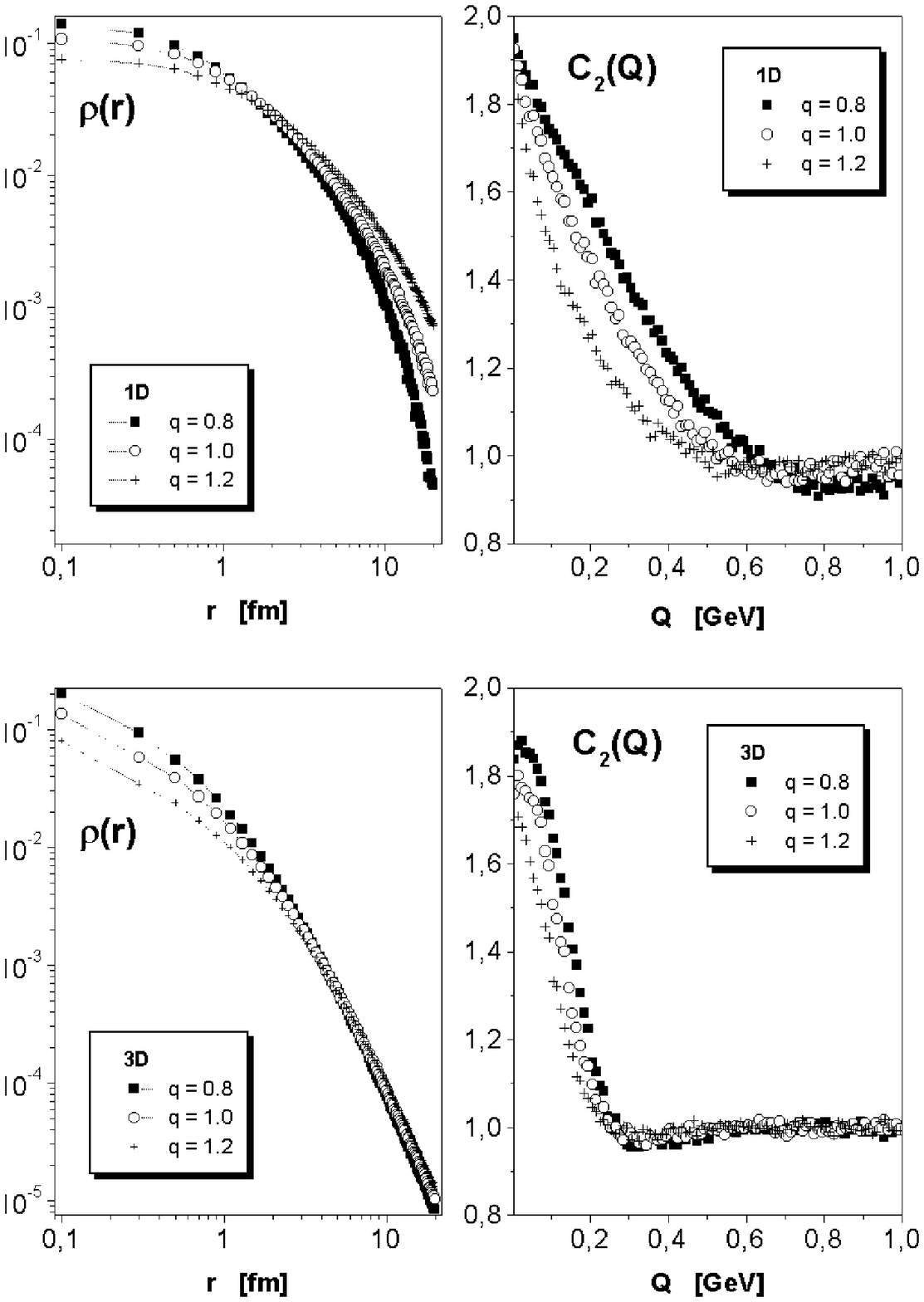}
\end{picture}
\end{figure}

\end{document}